\documentclass[twocolumn,american,prl,notitlepage,reprint,unsortedaddress,longbibliography]{revtex4-1}
\usepackage[T1]{fontenc}
\usepackage[latin9]{inputenc}
\usepackage{amsmath}
\usepackage{amssymb}
\usepackage{graphicx}
\usepackage{babel}
\begin{document}

\title{Stochastic Density Functional Theory at Finite Temperatures}

\author{Yael Cytter}

\affiliation{Fritz Haber Center for Molecular Dynamics and Institute of Chemistry,
The Hebrew University of Jerusalem, Jerusalem 9190401, Israel}

\author{Eran Rabani}
\email{eran.rabani@berkeley.edu}

\affiliation{Department of Chemistry, University of California and Lawrence Berkeley
National Laboratory, Berkeley, California 94720, U.S.A.}

\affiliation{The Raymond and Beverly Sackler Center for Computational Molecular
and Materials Science, Tel Aviv University, Tel Aviv, Israel 69978}

\author{Daniel Neuhauser}
\email{dxn@chem.ucla.edu}

\affiliation{Department of Chemistry, University of California at Los Angeles,
CA-90095 USA}

\author{Roi Baer}
\email{roi.baer@huji.ac.il}

\affiliation{Fritz Haber Center for Molecular Dynamics and Institute of Chemistry,
The Hebrew University of Jerusalem, Jerusalem 9190401, Israel}
\begin{abstract}
Simulations in the warm dense matter regime using finite temperature
Kohn-Sham density functional theory (FT-KS-DFT), while frequently
used, are computationally expensive due to the partial occupation
of a very large number of high-energy KS eigenstates which are obtained
from subspace diagonalization. We have developed a stochastic method
for applying FT-KS-DFT, that overcomes the bottleneck of calculating
the occupied KS orbitals by directly obtaining the density from the
KS Hamiltonian. The proposed algorithm, scales as $O\left(NT^{-1}\right)$
and is compared with the high-temperature limit scaling $O\left(N^{3}T^{3}\right)$
of the deterministic approach, where $N$ is the system size (number
of electrons, volume etc.) and $T$ is the temperature. The method
has been implemented in a plane-waves code within the local density
approximation (LDA); we demonstrate its efficiency, statistical errors
and bias in the estimation of the free energy per electron for a diamond
structure silicon. The bias is small compared to the fluctuations,
and is independent of system size. In addition to calculating the
free energy itself, one can also use the method to calculate its derivatives
and obtain the equations of state.

\end{abstract}
\maketitle

\section{Introduction}

Electronic structure calculations coupled with molecular dynamics
trajectory sampling, form a reliable and important source of information
concerning the properties materials at the warm dense matter (WDM)
regime. The main challenges lie in the determination of the equation
of state (EOS) of such systems~\cite{clerouin2005electrical,Kowalski2007,nettelmann2008ab,Saiz2008,faussurier2010equation,Morales2010},
addressing their various phase transition boundaries~\cite{Ernstorfer2009,Knudson2015},
and predicting shock-wave propagation characteristics, ~\cite{Militzer2001,Dharma-Wardana2006,mintsev2015transport}
as well as their transport and optical properties~\cite{Nagler2009,vlcek2012electrical,clerouin2012database,Pan2013,Apfelbaum2013,Sato2014}.

Reliable and predictive computational approaches should be based on
\emph{ab initio }calculations, and these usually fall within the Green's
function methods (GF)~\cite{Faleev2006,Kananenka2016,Neuhauser2017,Kas2017},
Monte-Carlo (MC) techniques~\cite{Militzer2001,Driver2012,Brown2014}
and density functional theory (DFT) ~\cite{Hohenberg1964,Mermin1965,pribram2014thermal}.
The fact that GF and MC methods are expensive is exacerbated by the
need to repeat the electronic calculation for the many nuclear configurations
along a molecular dynamics trajectory. 

Among the ab initio approaches, DFT methods emerge as an ideal framework,
combining useful accuracy and applicability. We differentiate between
orbital-free DFT~\cite{Wang1998,cangi2015efficient,Karasiev2012,karasiev2014accurate,Karasiev2015,Sjostrom2014}
and finite-temperature Kohn-Sham (FT-KS) approaches~\cite{Kohn1965,Pittalis2011}.
The former involves very moderate computational effort but is of limited
accuracy due to the use of approximate kinetic energy and entropy
functional. The latter class of DFT approaches on the other hand yields
reliable and accurate results and is emerging as the method of choice
in the field, with applications rangingfrom short pulse laser simulations~\cite{Silvestrelli1996,Kietzmann2008}
and x-ray scattering~\cite{Plagemann2012,Baczewski2016} to properties
of astrophysical bodies~\cite{Bethkenhagen2017,Alavi1995,Schottler2016,Knudson2015}.

The benefits of using the FT-KS method stems from the mapping of the
interacting system onto the non-interacting one governed by the single-particle
KS Hamiltonian. This, however, comes with a price, since in finite
temperature the eigenstates of the KS Hamiltonian are all formally
occupied according to the Fermi-Dirac distribution, thus requiring
the calculation of all the non-negligibly occupied eigenstates ($N_{occ}$
). This numerical task scales typically as $O\left(N_{occ}^{2}N\right)$,
where $N$ indicates the system size (volume, number of electrons,
etc). From the entropy of the non-interacting homogeneous electron
gas we find that the proportionality of the number of occupied states
with temperature is $N_{occ}\propto NT^{3/2}$, where $T$ is the
temperature (In this model the entropy is also proportional to a term
that is independent of $T$, so that in $T=0$ the number of occupied
states goes to a constant. However, one should note that the model
only holds in the limit of high temperatures). Consequently, the CPU
time of a FT-KS-DFT calculation is expected to scale as $O\left(T^{3}N^{3}\right)$,
which rises rapidly with temperatures and densities of the system.
Moreover, it grows significantly when the system is near a phase transition, where physical length-scales are large. 
The purpose of this paper is to propose an alternative implementation
of the FT-KS-DFT, based on a stochastic approach~\cite{Baer2013},
in which the CPU time increases linearly with system size and inverse
temperature. The scaling of the proposed method is therefore $O\left(NT^{-1}\right)$
to be compared with the $O\left(T^{3}N^{3}\right)$ scaling of conventional
FT-KS-DFT. Stochastic methods for electronic structure have been developed
recently and has shown to be highly efficacious in lowering the algorithmic
complexity of a variety of electronic structure calculations~\cite{Neuhauser2014,Neuhauser2013a,Neuhauser2013,Neuhauser2014a,Neuhauser2015,Neuhauser2017,arnon2017equilibrium,Rabani2015,takeshita2017stochastic}.
This paper shows that such an approach can also be useful for research
done in WDM and related fields. We focus on the electronic structure aspect of the free energy, neglecting the contribution of the nuclear kinetic energy and entropy, both of which requires an additional effort (such as a molecular dynamics sampling technique). In anticipation of the disorder created by the molecular dynamics sampling, we do not exploit the symmetry of the ordered lattice (as often done by k-point sampling techniques) which we use to demonstrate the method.

We present the method in Section~\ref{sec:Method}, and study its
validity by examining the convergence to the known (deterministic)
free energy and the stochastic noise accompanying the calculations
in Section~\ref{sec:Convergence-and-statistical}. In Section~\ref{sec:Electronic-equation-of}
we show how equations of state can be computed in the presence of
stochastic noise.

\section{\label{sec:Method}Method}

\subsection{FT-KS-DFT formalism}

Consider an ensemble of interacting electrons in inverse temperature
$\beta=\frac{1}{k_{B}T}$ and chemical potential $\mu$. The grand
canonical potential operator describing the system would then be
\begin{equation}
\hat{\Omega}=\hat{H}-T\hat{S}-\mu\hat{N}
\end{equation}
where $N$ is the number of electrons, $\hat{S}$ is the entropy and $\hat{H}$
is the interacting Hamiltonian, defined as
\[
\hat{H}=-\frac{1}{2m_{e}}\nabla^{2}+\hat{v}_{ee}+\hat{v}_{ext}\;;
\]
here $m_{e}$ is the electron's mass, $\hbar$ is Planck's constant,
$v_{ee}$ represents the interaction between the electrons and $\hat{v}_{ext}$
is the potential of interaction between the electrons and the nuclei
as well as other external fields.

The FT-KS-DFT method maps the interacting system onto an ensemble
of non-interacting electrons, the KS system, with the same one-electron
density $n\left(\boldsymbol{r}\right)$ and, commonly though not compulsory,
the same inverse temperature $\beta$ and chemical potential $\mu$
\cite{Parr1989}. These non-interacting electrons are described by
the KS Hamiltonian
\begin{equation}
\hat{h}=-\frac{\hbar^{2}}{2m_{e}}\nabla^{2}+v_{KS}\left(\boldsymbol{r}\right)\;.\label{eq:hKS}
\end{equation}
The potential $v_{KS}\left(\boldsymbol{r}\right)$ is given by

\begin{equation}
v_{KS}\left[n\right]\left(\boldsymbol{r}\right)=v_{ext}\left(\boldsymbol{r}\right)+v_{H}\left[n\right]\left(\boldsymbol{r}\right)+v_{xc}\left[n\right]\left(\boldsymbol{r}\right),\label{eq:vKS}
\end{equation}
where 
\begin{equation}
v_{H}\left[n\right]\left(\boldsymbol{r}\right)=\int n\left(\boldsymbol{r}'\right)\left|\boldsymbol{r}-\boldsymbol{r}'\right|^{-1}d^{3}r'
\end{equation}
is the Hartree potential and $v_{xc}\left[n\right]\left(\boldsymbol{r}\right)=\frac{\delta\Omega_{xc}\left[n\right]}{\delta n\left(\boldsymbol{r}\right)}$
is the exchange-correlation potential, which is a functional derivative
of the exchange-correlation grand canonical potential $\Omega_{xc}$.
This exchange correlation functional includes the differences between
the interacting and non-interacting system's kinetic energy and entropy,
as well as the difference between the full Coulomb repulsion energy
and the Hartree energy, defined as 
\begin{equation}
E_{H}=\frac{1}{2}\iint\frac{n\left(\boldsymbol{r}\right)n\left(\boldsymbol{r}'\right)}{\left|\boldsymbol{r}-\boldsymbol{r'}\right|}d^{3}r'd^{3}r.
\end{equation}
 The method can be useful if an efficacious approximation to $\Omega_{xc}\left[n\right]$
is available. The local density approximation provides such an approximate
functional~\cite{Kohn1965}:
\begin{align}
\Omega_{xc}^{LDA}\left[n;\beta\right] & =\int\omega_{xc}\left(n\left(\boldsymbol{r}\right),\beta\right)n\left(\boldsymbol{r}\right)d^{3}r\\
v_{xc}^{LDA}\left[n;\beta\right]\left(\boldsymbol{r}\right) & =\omega_{xc}\left(n\left(\boldsymbol{r}\right);\beta\right)+\omega_{xc}^{\prime}\left(n\left(\boldsymbol{r}\right);\beta\right)n\left(\boldsymbol{r}\right)\label{vxcLDA(r)}
\end{align}
where $\omega_{xc}\left(n;\beta\right)$ is the exchange-correlation
free energy per electron for a homogeneous electron gas at density
$n$ and inverse temperature $\beta$, parameterized based on Monte
Carlo free energy calculations~\cite{karasiev2014accurate}.

The system's electronic density is given as 
\begin{align}
n(\boldsymbol{r}) & =\sum_{i}f_{\beta,\mu}\left(\varepsilon_{i}\right)|\phi_{i}\left(\boldsymbol{r}\right)|^{2}\label{eq:density}\\
 & =tr\left[f_{\beta,\mu}\left(\hat{h}\right)\hat{n}(\boldsymbol{r})\right]\label{eq:density-as-trace}
\end{align}
where $\hat{n}\left(\boldsymbol{r}\right)=\delta\left(\hat{\boldsymbol{r}}-\boldsymbol{r}\right)$
is the electron density operator ($\hat{\boldsymbol{r}}$ is the position of the electron),
\begin{equation}
f_{\beta,\mu}\left(\varepsilon\right)=\text{\ensuremath{\frac{1}{1+e^{\beta\left(\varepsilon-\mu\right)}}}}\label{eq:fermi-dirac}
\end{equation}
is the Fermi-Dirac distribution and $\phi_i(\boldsymbol{r})$ ($\varepsilon_i$) is the eigenfunction (eigenvalue) of the self-consistent KS Hamiltonian:

\begin{equation}
\hat{h}\phi_{i}\left(\boldsymbol{r}\right)=\varepsilon_{i}\phi_{i}\left(\boldsymbol{r}\right).\label{eq:KS_hamiltonian}
\end{equation}

We then construct $v_{KS}$ according
to Eq.$\left(\ref{eq:vKS}\right)$ in order to solve Eq.$\left(\ref{eq:KS_hamiltonian}\right)$
again. The procedure is repeated until convergence is achieved. 

Once the density is obtained, the grand canonical free energy of the
interacting system, when the nuclear kinetic energy is neglected \cite{Comment},
is given as: 
\begin{equation}
\Omega=\Omega_{KS}-E_{H}\left[n\right]-\int n\left(\boldsymbol{r}\right)v_{xc}\left(\boldsymbol{r}\right)d^{3}r+\Omega_{xc}\left[n\right]+E_{N}\label{eq:G}
\end{equation}
where $E_{N}$ is the (classical) nuclear-nuclear repulsion energy
and 
\begin{align}
\Omega_{KS} & =E_{KS}-\mu N-TS_{KS}\label{eq:GKS}
\end{align}
is the grand canonical potential of the non-interacting system. Here,
\begin{equation}
E_{KS}=tr\left[f_{\beta,\mu}\left(\hat{h}\right)\hat{h}\right]\;.\label{eq:Eks-trace}
\end{equation}
 Furthermore,
\begin{equation}
S_{KS}=k_{B}tr\left[f_{\beta,\mu}\left(\hat{h}\right)\ln f_{\beta,\mu}\left(\hat{h}\right)+\bar{f}_{\beta,\mu}\left(\hat{h}\right)\ln\bar{f}_{\beta,\mu}\left(\hat{h}\right)\right]\label{eq:Sks-trace}
\end{equation}
is the entropy of the non-interacting electrons where we use the notation
$\bar{f}\equiv1-f$, and 
\begin{equation}
N_{e}=\int n\left(\boldsymbol{r}\right)d^{3}r=tr\left[f_{\beta,\mu}\left(\hat{h}\right)\right]\;.\label{eq:Ne-trace}
\end{equation}

As can be seen in Eqs.(\ref{eq:density-as-trace}) and (\ref{eq:Eks-trace})-(\ref{eq:Ne-trace})
all the quantities above can be expressed as traces. The series of
iterations involved in the FT-KS-DFT method can then be described
in the following manner: a previous guess density $n^{prev}\left(\boldsymbol{r}\right)$
is used to construct a KS potential $v_{KS}\left[n^{prev}\right]\left(\boldsymbol{r}\right)$,
from which a new guess density $n^{new}\left(\boldsymbol{r}\right)$
is obtained: 
\begin{align}
n^{prev} & \underset{\text{Eq.}\,(\ref{eq:vKS})}{\longrightarrow}v_{KS}\left[n^{prev}\right]\to\hat{h}\underset{\text{Eq.}\,(\ref{eq:density-as-trace})}{\longrightarrow}n^{new}\left(\boldsymbol{r}\right)\;.\label{eq:SCF}
\end{align}
These iterations are repeated until the previous and new densities
are equal to one another, in which case a self-consistent-field (SCF)
density $n\left(\boldsymbol{r}\right)$ is obtained, so that:
\begin{equation}
n^{prev}=n^{new}\equiv n\;.\label{eq:converged}
\end{equation}

\subsection{\label{subsec:Stochastic-approach-to}Stochastic approach to FT-KS-DFT}

The stochastic approach to FT-KS-DFT (sFT-KS-DFT) exploits the fact
that all terms in the free energy of Eq.~$\left(\ref{eq:GKS}\right)$
are expressed using traces over appropriate operators. These traces
are then estimated by the stochastic trace formula~\cite{Hutchinson1990}:
\begin{equation}
tr\left[\hat{A}\right]=\text{E}\left\{ \left\langle \chi\left|\hat{A}\right|\chi\right\rangle \right\} ,
\end{equation}
where $\hat{A}$ is an arbitrary operator, $\chi$ is a random ket
and $\text{E}\left\{ \cdots\right\} $ is the statistical average
value of the random variable appearing inside the curly brackets.
If we use a Cartesian grid of $N_{g}$ grid points $\boldsymbol{r}$
to represent wave functions and operators in real-space then the ket
$\chi$ is a random orbital and at each grid point $\chi\left(\boldsymbol{r}\right)$
is a random variable with zero mean, $\text{E}\left[\left\{ \chi\left(\boldsymbol{r}\right)\right\} \right]=0$
, and a covariance given by
\begin{equation}
\text{E}\left\{ \chi\left(\boldsymbol{r}\right)\chi\left(\boldsymbol{r}'\right)^{*}\right\} =\delta^{-3}\delta_{\boldsymbol{r}\boldsymbol{r}'}
\end{equation}
where $\delta$ is the grid spacing. This requirement on the random
orbital can be achieved by choosing $\chi\left(\boldsymbol{r}\right)=\delta^{-3/2}e^{i\theta\left(\boldsymbol{r}\right)}$
for each grid point $\boldsymbol{r}$, where $\theta\left(\boldsymbol{r}\right)$
is an independent random number in the $\left[0,2\pi\right]$ interval. 

Assuming the Hamiltonian $\hat{h}$ is known, the FT density can be
computed from the trace formula, using $I$ stochastic orbitals $\chi_{i}$,
$i=1,\dots,I$ as follows

\begin{equation}
n_{I}\left(\boldsymbol{r}\right)=\frac{1}{I}\sum_{i=1}^{I}\left|\xi_{i}\left(\boldsymbol{r}\right)\right|^{2}\label{eq:n(r).eq.xi(r)^2}
\end{equation}
where $\xi_{i}\left(\boldsymbol{r}\right)$ is a thermally filtered
random orbital, given by:
\begin{equation}
\xi_{i}\left(\boldsymbol{r}\right)=\left\langle \boldsymbol{r}\right|\theta_{\beta,\mu}\left(\hat{h}\right)\left|\chi_{i}\right\rangle \label{eq:xi(r)}
\end{equation}
where $\theta_{\beta,\mu}\left(\varepsilon\right)=\sqrt{f_{\beta,\mu}\left(\varepsilon\right)}$. 

The SCF procedure in the stochastic approach involves a previous stochastic
guess density $n_{I}^{prev}\left(\boldsymbol{r}\right)$ and the following
process to update it, analogous to Eq.~$\left(\ref{eq:SCF}\right)$:
\begin{align}
n_{I}^{prev} & \underset{\text{Eq.}\,(\ref{eq:vKS})}{\longrightarrow}v_{KS}\left[n_{I}^{prev}\right]\to\hat{h}\underset{\text{Eq.}\,(\ref{eq:n(r).eq.xi(r)^2})}{\longrightarrow}n_{I}^{new}\left(\boldsymbol{r}\right).\label{eq:sSCF}
\end{align}
These iterations are repeated until the previous and new densities
are equal, in which case a self-consistent-field (SCF) density $n_{I}\left(\boldsymbol{r}\right)$
is obtained: 
\begin{equation}
n_{I}^{prev}=n_{I}^{new}\equiv n_{I}.\label{eq:converged-I}
\end{equation}
The random density $n_{I}$ is distributed with a mean $\text{E}\left\{ n_{I}\right\} $
and a certain \emph{standard deviation }proportional to $I^{-1/2}$
representing the statistical error. A second part of the statistical
error is the \emph{bias}, defined as
\begin{equation}
bias=\text{E}\left\{ n_{I}\right\} -n.
\end{equation}
The origin of the bias is the nonlinear nature of the SCF cycle in
Eq.~$\left(\ref{eq:SCF}\right)$ and this error can be shown to diminish
asymptotically linearly with $I^{-1}$ (for further reading see Ref.~\citenum{Bohm2010});
we will see that is indeed the case in actual calculations presented
below. The general conclusion is, that as $I$ increases the bias
diminishes \emph{faster }than the standard deviation.

\subsection{The Chebyshev expansion\label{subsec:The-Chebyshev-expansion}}

For each $\chi\left(\boldsymbol{r}\right)$, the calculation of $\xi\left(\boldsymbol{r}\right)$
employs a Chebyshev polynomial expansion, i.e. 
\begin{equation}
\xi\left(\boldsymbol{r}\right)=\sum_{n=0}^{N_{C}-1}C_{n}\left[\theta_{\beta,\mu}\right]\phi_{n}\left(\boldsymbol{r}\right)\label{eq:ChebyshevExp}
\end{equation}
where $\phi_{0}\left(\boldsymbol{r}\right)=\chi\left(\boldsymbol{r}\right)$,
$\phi_{1}\left(\boldsymbol{r}\right)=\hat{h}_{N}\chi\left(\boldsymbol{r}\right)$,
and for $n>1$, $\phi_{n}\left(\boldsymbol{r}\right)=2\hat{h}_{N}\phi_{n-1}\left(\boldsymbol{r}\right)-\phi_{n-2}\left(\boldsymbol{r}\right)$,
with the normalized Hamiltonian 
\begin{equation}
\hat{h}_{N}=\frac{\hat{h}-\frac{1}{2}\left(E_{max}+E_{min}\right)}{\frac{1}{2}\left(E_{max}-E_{min}\right)}=\frac{\hat{h}-\bar{E}}{\Delta E}
\end{equation}
where $E_{max}$ ($E_{min}$) is an upper bound on the maximal (lower
bound on the minimal) eigenvalue of $\hat{h}$. The coefficients $C_{n}\left[F\right]$
are the Chebyshev coefficients corresponding to a function $F\left(\varepsilon\right)$
(which is equal to $\theta_{\beta,\mu}$ in this case). They are given
by~\cite{Press2007}:
\begin{equation}
C_{n}=\frac{2-\delta_{n0}}{2N_{C}}e^{i\frac{n\pi}{2N_{C}}}\tilde{F}_{n},\,\,n=0,1,\dots,N_{C}-1\label{eq:ChebyshevCoeffs}
\end{equation}
where $\tilde{F_{n}}$ are the first $N_{C}$ terms of the fast Fourier
transform of the series 
\begin{equation}
F_{k}=F\left(x_{k}\Delta E+\bar{E}\right),\;k=0,1,\dots,2N_{C}-1
\end{equation}
where $x_{k}=\cos\frac{\pi\left(k+\frac{1}{2}\right)}{N_{C}}$. In
the case of Eq.~(\ref{eq:ChebyshevExp}) we take this function as
$F\left(\varepsilon\right)=\theta_{\beta\mu}\left(\varepsilon\right)$.
The expansion length $N_{C}$ is selected such that $\left|C_{n}\right|<10^{-9}$
for $n>N_{C}$. I can typically be shown that~\cite{Baer1997b}: 

\begin{equation}
N_{C}\propto\beta\Delta E.\label{eq:NC=00003Dbeta*DE}
\end{equation}
Since $\Delta E$ is half the difference between $E_{max}$ and $E_{min}$
where $E_{max}$ is usually determined by the kinetic energy cutoff
and $E_{min}$ is determined by the ground state screened potential
(including the non-local part of the pseudopotential), the number
of terms in Chebyshev expansion is largely independent of system size.

\subsection{Chebyshev moments}

Besides the density, other quantities of interest (see Eqs.~(\ref{eq:Eks-trace}),
(\ref{eq:Sks-trace}) and (\ref{eq:Ne-trace})) are all traces of
certain functions $F\left(\hat{h}\right)$ of the KS Hamiltonian $\hat{h}$.
The calculation of these quantities can all be expressed as sums of
the form:
\begin{equation}
tr\left[F\left(\hat{h}\right)\right]=\sum_{n=0}^{N_{C}}C_{n}\left[F\right]M_{n}\label{eq:trace=00005Bf(hN)=00005D}
\end{equation}
where $C_{n}\left[F\right]$ are the Chebyshev expansion coefficients
defined in Eq.~(\ref{eq:ChebyshevCoeffs}) and 
\begin{equation}
M_{n}=\text{E}\left\{ \left\langle \left.\chi\right|\phi_{n}\right\rangle \right\} ,
\end{equation}
are the Chebyshev \emph{moments}~\cite{Wang1994b}. 

All the results shown in this paper are within the canonical ensemble,
having the Helmholtz free energy 
\begin{align}
A\left(\beta,V,N_{e}\right) & =\Omega\left(\beta,\mu,V\right)+N_{e}\mu\;.
\end{align}
 To obtain a constant number of electrons $N_{e}$ at each iteration,
we include a step of search for the value of $\mu$ for which the
number of electrons, defined in Eq.$\left(\ref{eq:Ne-trace}\right)$
and calculated using the moments, is equal to the desired number of
electrons. The free energy is then
\begin{equation}
A\left(\beta,V,N_{e}\right)=E_{KS}\left(\beta,V,\mu\left(N_{e}\right)\right)-TS_{KS}\left(\beta,\mu\left(N_{e}\right)\right)+E_{N}.
\end{equation}

The main advantage of the sFT-KS-DFT method is its lower scaling.
In the FT-KS-DFT calculation the density is represented as a sum of the
square absolute value Kohn-Sham eigenfunctions  (each multiplied by its electronic occupation). To calculate these on a grid,
one needs to invest $O\left(N_{occ}^{2}N_{g}\right)$ operations and
$O\left(N_{occ}N_{g}\right)$ memory capacity for storage. Since $N_{occ}$
increases quickly with $\beta^{-1}$ (see discussion in the introduction),
the FT-KS-DFT method is a very expensive way to study warm dense matter.
In contrast, sFT-KS-DFT requires $N_{C}$ applications of the KS Hamiltonian
$\hat{h}$, to a set of $I$ $\chi$'s, a step of order $N_{C}\times I\times N_{g}\ln N_{g}$
floating-point operations (the $\ln N_{g}$ is due to the fast Fourier
transform required for the kinetic energy operation).

\begin{figure}
\begin{centering}
\includegraphics[width=0.8\columnwidth]{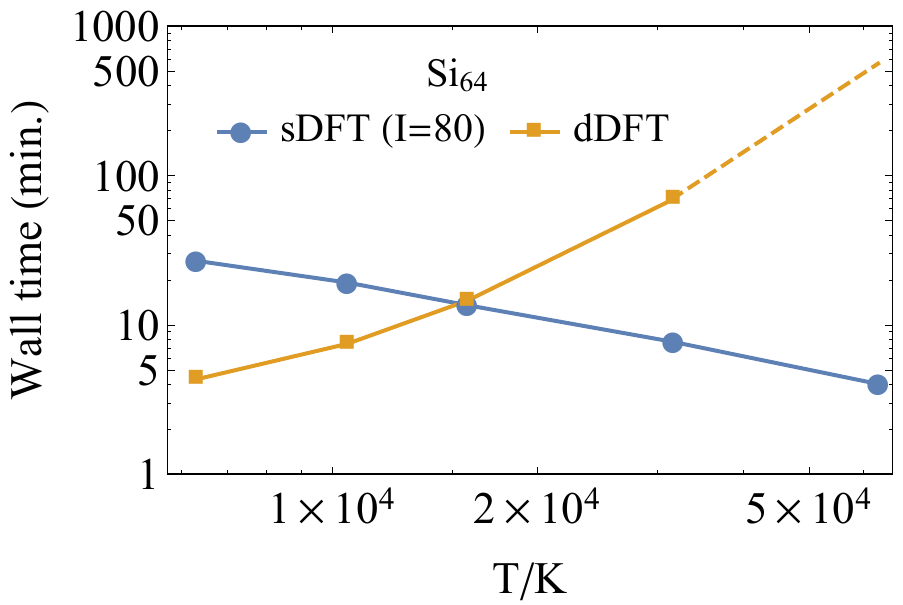}
\par\end{centering}
\caption{\label{fig:CPU_vs_Beta} CPU wall-time for self-consistent KS-DFT
calculations using the stochastic (sDFT, $I=80$ stochastic orbitals)
and deterministic (single thread dDFT, Quantum Espresso~\cite{Giannozzi2009})
calculations on $\text{Si}_{64}$ having a lattice constant of $21a_{0}$.
The dashed line is the expected $O\left(T^{3}\right)$ extrapolation
for the dDFT timings.}
\end{figure}

In Section~\ref{sec:Convergence-and-statistical} we demonstrate
how the error per electron, determined by $I$, does not increase
with system size. This is compatible with findings shown in previous
work for finite systems described in Ref. \cite{Baer2013}. In addition,
the prior work demonstrates that as explained in Subsection~\ref{subsec:The-Chebyshev-expansion},
$N_{C}$ at a given $\beta$ is independent of system size as well.
Therefore, since $I$ and $N_{C}$ do not increase with system size
the computational effort scales linearly. In addition, since the Chebyshev
expansion length $N_{C}$ is proportional to $\beta$ (see Eq.~(\ref{eq:NC=00003Dbeta*DE})),
the CPU time actually drops as $\beta$ decreases (temperature rises)
as can be seen in Fig.~\ref{fig:CPU_vs_Beta}. Moreover, since the
procedure is based on averaging over random values of the density,
the process of attaining the different values of $n_{i}\left(\boldsymbol{r}\right)$
can be done most naturally in parallel.

\section{\label{sec:Convergence-and-statistical}Results: Convergence and
statistical errors}

The stochastic method described in the previous Subsection has been
implemented within our \emph{Inbar} \footnote{\emph{Inbar }(Hebrew), \emph{anbar }(Arabic), \emph{amber }or \emph{\={e}lectrum}
(Latin) and \emph{electron }in Greek is the yellow-brownish fossil
resin of vegetable origin negatively charged by friction.} plane-waves DFT code and the resulting implementation is dubbed \emph{sInbar}.
For demonstrating the code, we use silicon in a FCC diamond structure
with periodic boundary conditions described by the ground state local
density approximation (LDA). We consider $\text{Si}_{8}$, $\text{Si}_{64}$,
$\text{Si}_{216}$ and $\text{Si}_{512}$ having respectively 32,
256, 864 and 2048 valence electrons, each using a cubic supercell
size of length $a$, $2a$, $3a$ and $4a$, and the Fourier grid
includes $N_{g}$, $2^{3}N_{g}$, $3^{3}N_{g}$ and $4^{3}N_{g}$
grid points respectively, where $N_{g}=30^{3}$. The kinetic energy
cutoff is 20Ry and Troullier-Martins norm-conserving pseudopotentials~\cite{Troullier1991}
within the Kleinman-Bylander representation~\cite{Kleinman1982}
are deployed for describing the electron-nucleus interactions. In
the temperature regime used here, based on the results of Ref.~\cite{karasiev2014accurate},
it is justified to use ground-state LDA, for which we adopt here the
parameterization of Ref.~\cite{Perdew1992a}.

\begin{figure*}
\begin{centering}
\includegraphics[width=0.32\textwidth]{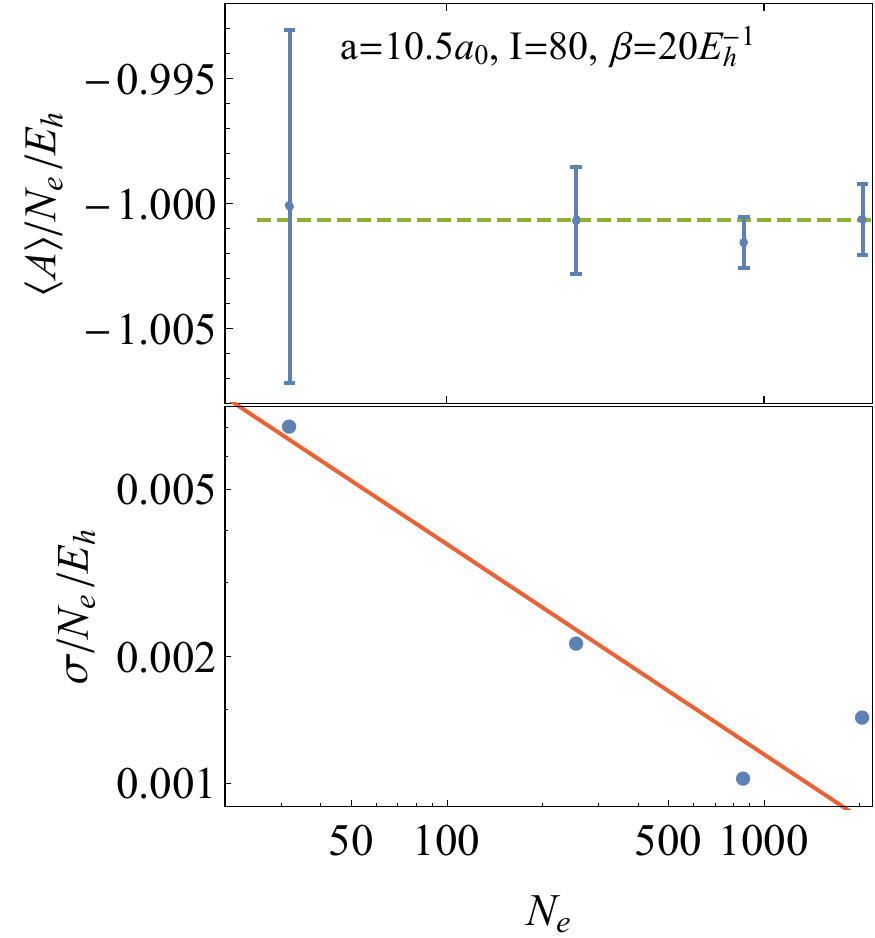}
\includegraphics[width=0.32\textwidth]{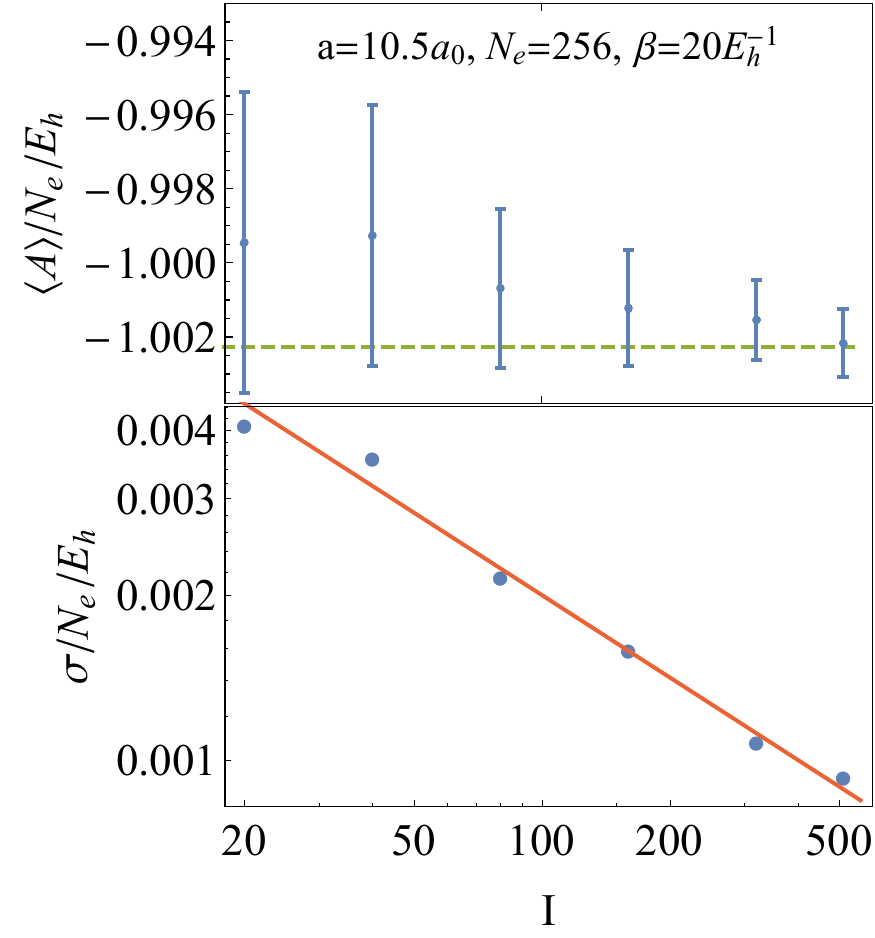}
\includegraphics[width=0.32\textwidth]{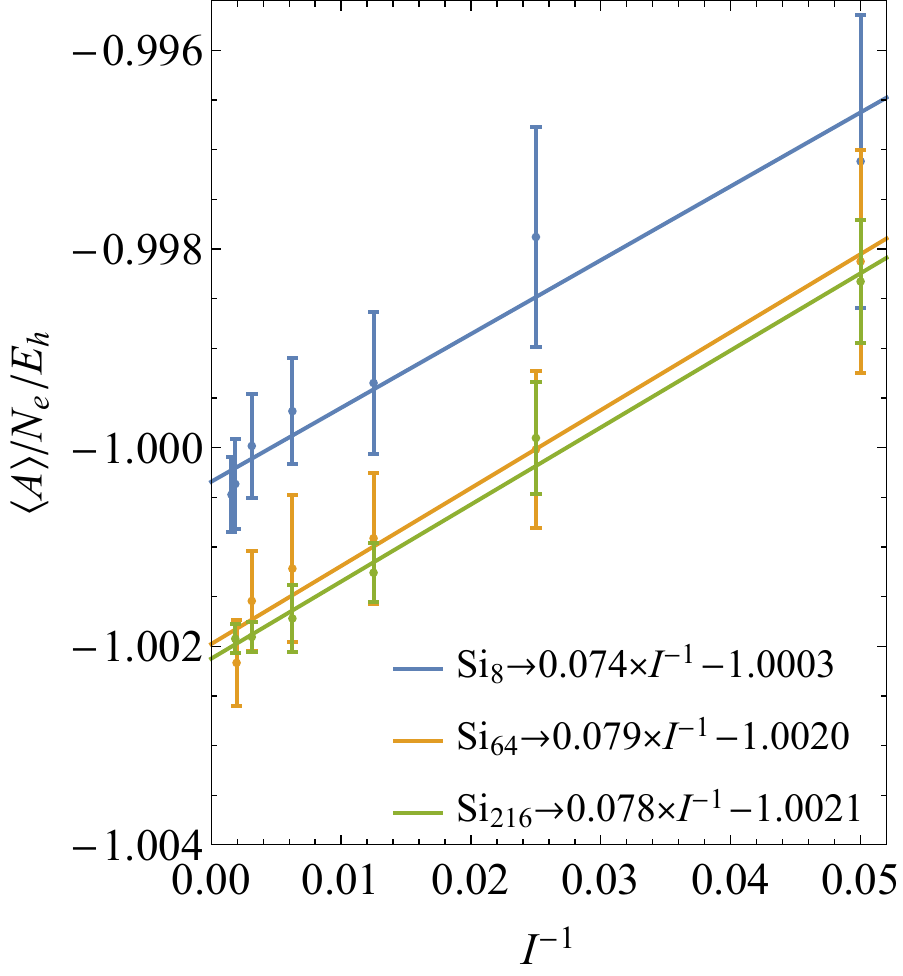}
\par\end{centering}
\caption{\label{fig:Statistical-errors}\textbf{Left panels}: We show in the
top left panel the estimated expected value of the Helmholtz free
energy per electron (dots) and its square-root-variance $\sigma$
(half length of error bars $\pm\sigma$ which is also shown in the
log-plot of the bottom panel) as a function of the number of electrons
$N_{e}$ in four Si supercell sizes (see text) using $I=80$ stochastic
orbitals. The dashed green line is a guide to the eye designating
the free energy per electron of the largest system. \textbf{Middle
panels}: in the top middle panel we show the estimated expected value
of the Helmholtz free energy per electron (dots) and its square-root-variance
$\sigma$ (which is also shown in the log plot of the bottom panel)
as a function of the number of stochastic orbitals $I$ for $\text{Si}_{64}$.
The dashed green line designates the deterministic value of the free
energy per electron for this system. \textbf{Right panel: }The 70\%
confidence intervals of the estimated Helmholtz free-energy per electron,
for several system sizes, as a function of the inverse number of stochastic
orbitals $I^{-1}$ the solid lines are best-fit linear curves for
the data (their equations are given in the legend). }
\end{figure*}

We now study the nature of the statistical errors in the Helmholtz
free energy estimation and their behavior as a function of sampling
and system size. In the left panel of Fig.~\ref{fig:Statistical-errors}
we show the statistics of the Helmholtz free energy per electron $\left\langle A\right\rangle /N_{e}$
estimates as a function of the number of electrons $N_{e}$ in the
unit cell using the four systems presented above at $\beta=20E_{h}^{-1}$,
keeping the number of stochastic orbitals fixed $I=80$ (we chose
this value for $I$ because results based on it are a good balance
between accuracy and computational effort for this system). For each
system we use 6 calculations to estimate the average and standard
deviation $\sigma/N_{e}$ presented in the figure. As shown in the
bottom left hand panel, as system size grows $\sigma/N_{e}$ drops
in proportion to $N_{e}^{-1/2}$ in accordance with the self averaging
effect~\cite{Baer2013}.

The effect of sample size is studied in the middle panel of Fig.~\ref{fig:Statistical-errors},
using the $\text{Si}_{64}$ system at $\beta=20E_{h}^{-1}$ for demonstration.
The standard deviation $\sigma/N_{e}$ decreases as the sample size
grows, roughly in proportion to $I^{-1/2}$ (as in the left panel,
we used 6 independent runs to estimate the mean and the error bars).
In addition to the statistical fluctuations the free energy estimate
$\left\langle A\right\rangle /N_{e}$ is seen to be biased towards
values larger than the deterministic value (dashed line). For $I>20$
the bias error is found to be smaller than the size of the fluctuation
$\sigma/N_{e}$ and as shown in the right panel of Fig.~\ref{fig:Statistical-errors}
the bias decreases linearly with $I^{-1}$ and thus diminishes faster
than the fluctuation as $I$ increases, as was also discussed in Section~\ref{subsec:Stochastic-approach-to}.
It is seen in the right panel of the figure that the bias error in
the free energy estimate $\left\langle A\right\rangle /N_{e}$ is
largely independent of system size. 

\begin{figure}
\includegraphics[width=0.8\columnwidth]{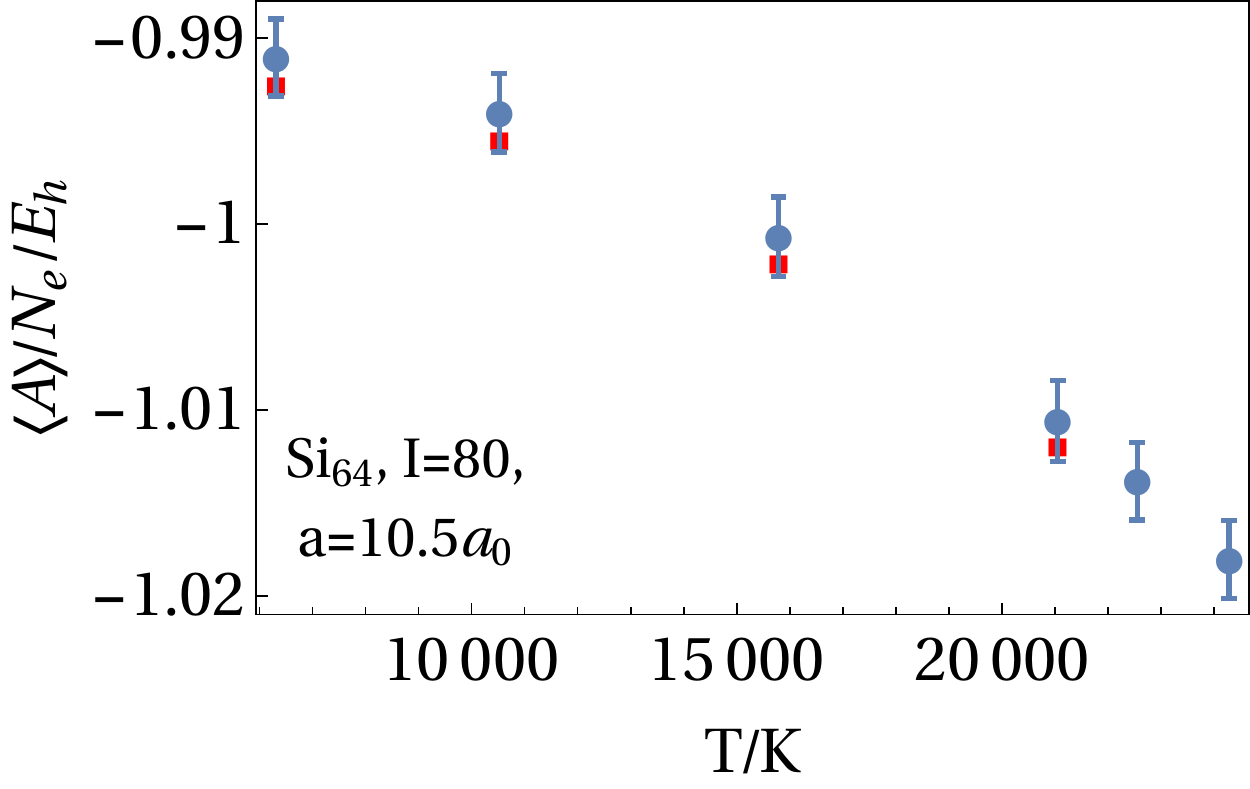}

\caption{\label{fig:AvsBeta}The estimated expected value of the Helmholtz
free energy per electron $\left\langle A\right\rangle /N_{e}$ (dots)
and its square-root-variance $\sigma$ (half length of error bars
$\pm\sigma$) for $\text{Si}_{64}$ as a function of inverse temperature
$\beta$ ,representing a single run of $I=80$ stochastic orbitals.
The expected value and standard deviation were estimated from six
independent such runs. The square symbols are the corresponding deterministic
values of the free energy.}
\end{figure}

The free energy as a function of temperature is shown in Fig.~\ref{fig:AvsBeta}
where one can see that the statistical fluctuation is not significantly
affected by the temperature. Deterministic results are also depicted
and once again, the expected values based on statistical estimates
are consistently above the deterministic results, showing a temperature-independent
statistical bias which is smaller than the fluctuation for $I=80$. 

\section{\label{sec:Electronic-equation-of}Equation of state calculations}

In order to calculate thermodynamic properties of the system one needs
to take derivatives of the free energy with respect to thermodynamic
variables, such as volume and pressure. In the present paper, as was
mentioned in Footnote~\cite{Comment}, we do not consider the free
energy resulting from the nuclear kinetic energy or entropy. As a
result, the equations of state we compute are mostly electronic, and
the nuclear positions simply comes in as the external potential alongside
the nuclear-nuclear repulsion term. Subsequently, we need to define,
perhaps arbitrarily, what change needs to be made when we change the
system volume. The most natural assumption is to preserve the FCC
diamond structure and impose cubic volume changes. The type of free
energy obtained from this calculation can be that of a system after
exposure to a short and powerful laser pulse, where due to a separation
of timescales the nuclei have not yet responded to the external field~\cite{Nagler2009,Ernstorfer2009}.

\begin{figure}
\begin{centering}
\includegraphics[width=0.8\columnwidth]{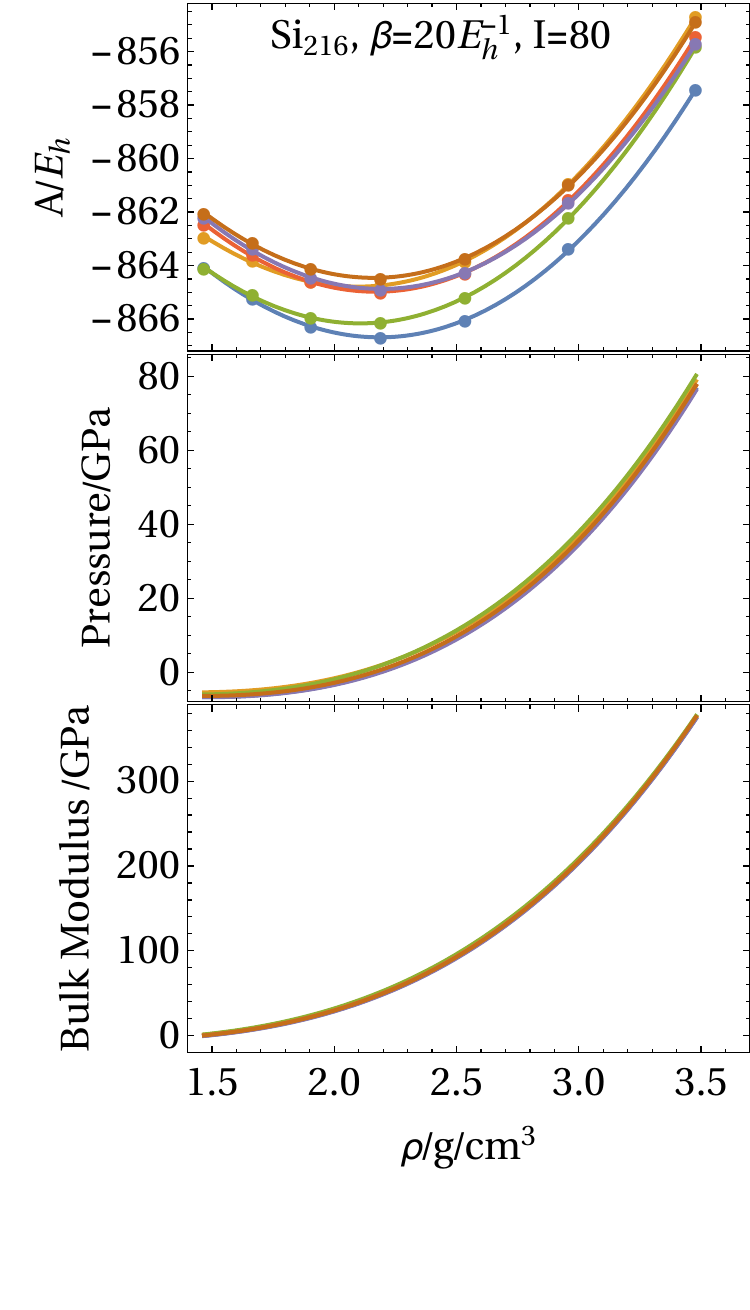}
\par\end{centering}
\caption{\label{fig:Aand Derivativesvsa}Top panel: The calculated values of
the Helmholtz free energy $A$ for the $\text{Si}_{216}$ system (with
$N_{e}=864$ electrons) at $\beta=20E_{h}^{-1}$, at several discrete
values of the density $\rho=N_{e}/V$ (shown as points on plot) calculated
for six independent seeds, i.e. each using six sets of $I=80$ stochastic
orbitals. The smooth lines express the free energy $A\left(\beta,N_{e},\rho\right)$
as a function of $\rho$ using cubic polynomials which best fit the
points. Middle and Bottom panels: the pressure and bulk modulus isotherms
derived from the six free energy curves of the top panel.}
\end{figure}

To address the practical problem of computing derivatives in the presence
of stochastic noise we calculate the free energy as a function of
a chosen parameter in a \emph{statistically correlated }way. For example,
we calculate the free Helmholtz energy $A\left(\beta,N_{e},\rho\right)$
for a given electron number $N_{e}$ and inverse temperature $\beta$
for several discrete values of the density $\rho=N_{e}/V$, where
$V=a^{3}$ is the cubic simulation cell volume and $a$ is its length.
This is done using the same number of Fourier grid points and the
same set of random phases for each stochastic orbital on the grid.
We demonstrate the results for the $\text{Si}_{216}$ system in Fig.~\ref{fig:Aand Derivativesvsa}
corresponding to six independent sets of $I=80$ stochastic orbitals
with which the sFT-KS-DFT calculations were performed. For each set
of free energies a 3rd degree polynomial is constructed to best-fit
the data. It is seen in the top panel of the figure that the free
energy data points are well-described by the polynomial. A higher
order fit does not significantly change the result shown here, indicative
of the low level of noise in each separate calculation. The statistical
fluctuations are evident in the slightly different shape and shift
of the polynomials. The derivatives of each of the free energy polynomials
can be used to calculate the corresponding pressure $P=-\left(\frac{\partial A}{\partial V}\right)_{\beta,N_{e}}$
and bulk modulus $B=-V\left(\frac{\partial P}{\partial V}\right)_{\beta,N}$,
both quantities are plotted, for each set of stochastic calculations,
in the middle and lower panels of Fig.~\ref{fig:Aand Derivativesvsa}.
The plot gives a sense of the behavior of the statistics of the derivatives
which seems well under control, showing that the equations of state
of the electrons are accessible using the stochastic approach. 

To get a more detailed description of the equation of state and its
statistical variance the above procedure is repeated using the same
sets of stochastic orbitals, for several values of $T$. In the top
panel of Fig.~\ref{fig:rho_BM_and_Cv_vs_T}, we show the isobar density
$\rho$ as a function of temperature for the $\text{Si}_{216}$ system,
for several values of the pressure $P$. The density decreases with
increasing temperature and pressure. The results of stochastic calculations
for $\text{Si}_{64}$ are shown as well (in darker colors) and a size
effect, where the density is too high in the small system, is noticeable
at high temperatures. We found that this high-temperature-low density
size-dependence is due mainly to the entropy term in the free energy. 

From the results shown in the figure, it is apparent that the standard
deviation in the calculations does not change as a function of temperature,
enabling good resolution with a clearly visible trend. We see that
regardless of the pressure, the density decreases with temperature,
i.e. the system expands. 

The bulk modulus shown in the bottom panel of Fig.~\ref{fig:rho_BM_and_Cv_vs_T}
seems to decrease as we go to higher temperatures, up until the point
where it is roughly estimated to vanish at $T\approx26,000K$, at
this point the density will be very low. The implication of these
results for fast electron heating by powerful lasers is that at short
time scales after the pulse, when nuclei are still cold, the material
can retain its elasticity even in temperatures of up to 20,000K. This
concept has been investigated both theoretically and experimentally
in relation to nonthermal melting \cite{SokolowskiTinten2003,Medvedev2015,Rousse2001},
where the potential energy surface changes as a result of excitation
of a large fraction of the valence electrons to the conductance band.
Previous work showed that neglecting electron-phonon interactions
leads to overestimation of the phase transition threshold in silicon
\cite{Medvedev2015}, a matter that could explain our results. In
this paper, however, our calculations are restricted to examination
of the breathing mode. To further explore the subject, molecular dynamics
has to be employed.

The heat capacity $C_{V}=-T\left(\frac{d^{2}A}{dT^{2}}\right)$ is
shown in Fig.~\ref{fig:rho_BM_and_Cv_vs_T} exhibiting low statistical
noise, that is hardly noticeable in the tested scales. We see that
the heat capacity grows almost linearly with temperature and goes
to zero as $T\rightarrow0$, in accordance with the third law of thermodynamics
applying for perfect crystals. To validate the calculation, we changed
the polynomial degree of the fit from 3rd to 5th order and saw almost
no difference in the heat capacity's behavior as a function of the
displayed temperatures. At higher temperatures, however, the fit becomes
more sensitive to the polynomial order, an effect amplified when looking
at its second derivative. This is a results of the steep decrease
of the free-energy as temperature increases, as seen in Fig. \ref{fig:AvsBeta}.
To avoid the inconsistency, at higher temperature range the free-energy
has to be samples more frequently. 
\begin{center}
\begin{figure}
\begin{centering}
\includegraphics[width=0.8\columnwidth]{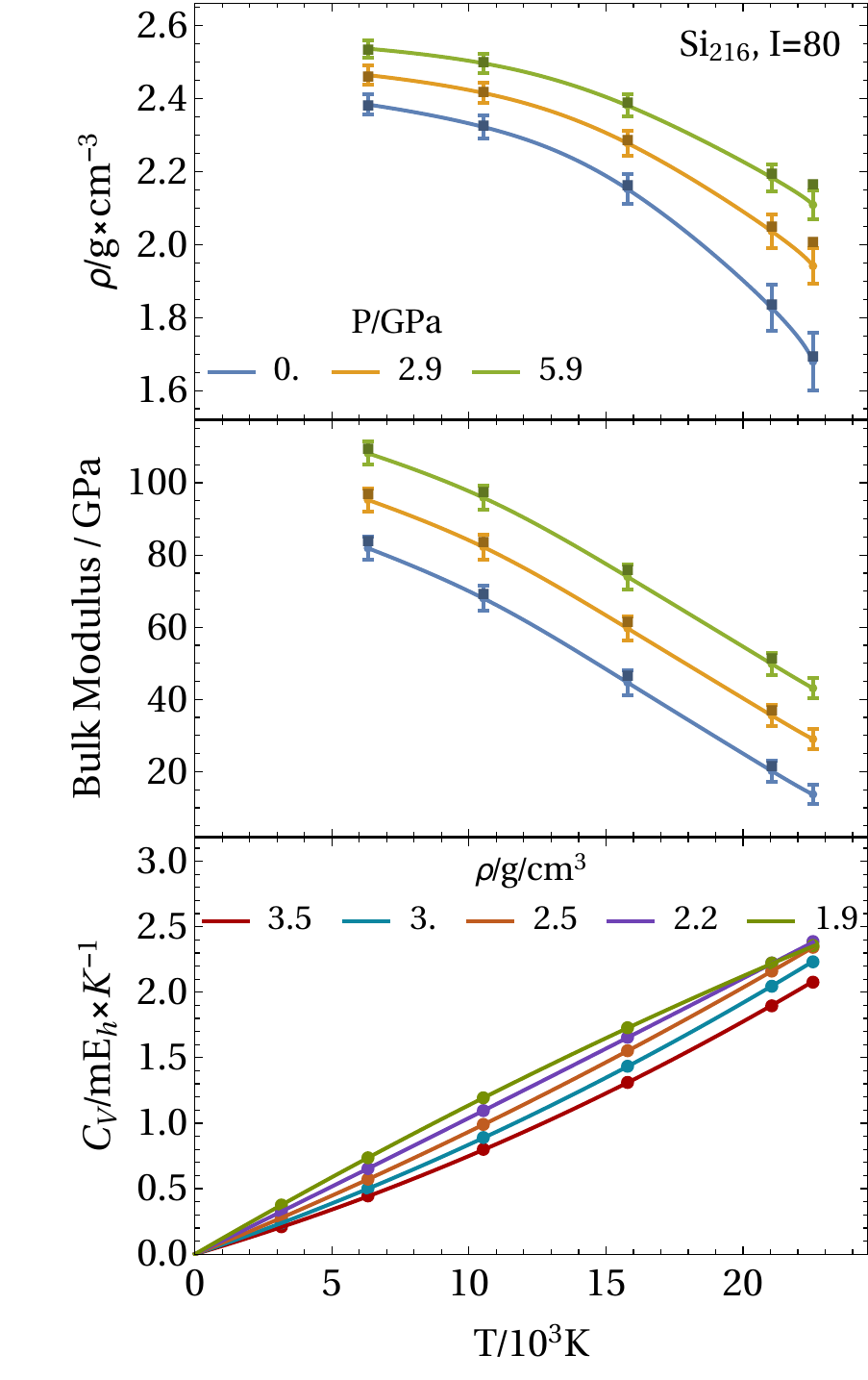}
\par\end{centering}
\centering{}\caption{\label{fig:rho_BM_and_Cv_vs_T}\textbf{Top two panels}: Isobars of
the density (\textbf{top panel}) and bulk modulus (\textbf{middle
panel}) in the $\text{Si}_{216}$ system, as function of temperature
under different pressures. The stochastic calculations were done using
$I=80$ stochastic orbitals and the data was discerned from the free
energy calculations discussed in the text. The darker squares are
the results of a deterministic calculation for the $\text{Si}_{64}$
system. \textbf{Bottom panel:} The heat capacity for several values
of the density as a function of temperature. The error bars are the
errors per one seed, and are the size of the markers. The lines are
the polynomial fits for a specific seed.}
\end{figure}
\par\end{center}

\section{Summary}

In this paper we introduced the stochastic approach for FT-KS-DFT
calculations to the warm dense matter regime. We analyzed the statistical
errors associated with the stochastic calculations and their dependence
on the number of iterations $I$, the system size and various parameters
such as the temperature and the density. We found that the fluctuations
in the estimates of the intensive properties decrease as $I^{-1/2}$
and as system size grows. The bias errors, resulting from the nonlinear
nature of the self consistent-field procedure, do not grow with system
size and decay as $I^{-1}$. In general, the bias errors turned out
to be small for the systems studied here. Furthermore, while both
errors do not depend on the temperature, calculation time is inversely
proportional to it, making the method highly efficient in the high
temperature regime. It has also been shown that the Helmholtz free
energy $A\left(\beta,\rho,N_{e}\right)$ can be computed as a smooth
and well-behaved function of its variables provided the same set of
stochastic orbitals are used. By exploiting this feature, we demonstrated
that the equations of state and the associated properties, such as
the pressure, heat capacity and bulk modulus become accessible as
derivatives. Our calculations did not use the symmetry properties
that allow efficient k-point sampling to be utilized, in anticipation
of the realistic cases where high temperature is associated with disorder
and non-symmetry. 

Future work in the field will include an implementation to molecular
dynamics sampling of the nuclear properties. Such an approach has
recently been shown viable at low temperatures using embedded fragments
which lower the statistical errors~\cite{arnon2017equilibrium}.
An additional future direction will examine the use of potential functional
theory \cite{Cangi2011,cangi2015efficient} for WDM calculations,
where we will study the possibility of lowering the variance by using
a coupling constant integration instead of a trace over the kinetic
energy operator.
\begin{acknowledgments}
We gratefully acknowledge the support of the NSF-ISF Israel-USA Binational
Science Foundation grant Number 2015687.
\end{acknowledgments}

\end{document}